\documentclass[useAMS,usenatbib]{mn2e}
\usepackage{epsfig}
\usepackage{pgfplots}
\usepackage{graphicx}
\usepackage{lscape}
\usepackage{amsmath}
\usepackage{amsbsy} 
\usepackage{rotating}
\usepackage{booktabs}
\usepackage{array}
\usepackage{lscape}
\usepackage{float}

\title[The thoroughly mixed Ib SN 1999dn]{Type Ib SN 1999dn as an example of the thoroughly mixed ejecta of Ib supernovae}

\author[Cano, Maeda \& Schulze]{\noindent Zach Cano$^{1}$\thanks{zcano@mail.com}, Keiichi Maeda$^{2,3}$ \& Steve Schulze$^{4,5}$ \\
\noindent $^1$Centre for Astrophysics and Cosmology, Science Institute, University of Iceland, Reykjavik, Iceland, 107.\\
$^2$Department of Astronomy, Kyoto University, Kitashirakawa-Oiwake-cho, Sakyo-ku, Kyoto 606-8502, Japan.\\
$^3$Kavli Institute for the Physics and Mathematics of the Universe (WPI), Todai Institutes for Advanced Study, The University of Tokyo,\\ 5-1-5 Kashiwanoha, Kashiwa, Chiba 277-8583, Japan.\\
$^4$Instituto de Astrof\'{i}sica, Facultad de F\'{i}sica, Pontificia Universidad Cat\'{o}lica de Chile, Casilla 306, Santiago 22, Chile. \\
$^5$Millennium Center for Supernova Science.\\
}

\begin{document}

\date{Accepted xx. Received xx; in original form xx}

\pagerange{\pageref{firstpage}--\pageref{lastpage}} \pubyear{2013}

\maketitle

\label{firstpage}

\begin{abstract}

We present the results of modelling archival observations of type Ib SN 1999dn.  In the spectra, two He I absorption features are seen: a slower component with larger opacity, and a more rapid He I component with smaller opacity.  Complementary results are obtained from modelling the bolometric light curve of SN 1999dn, where a two-zone model (dense inner region, and less dense outer region) provides a much better fit than a one-zone model.  A key result we find is that roughly equal amounts of radioactive material are found in both regions.  The two-zone analytical model provides a more realistic representation of the structure of the ejecta, including mixing and asymmetries, which offers a physical explanation for how the radioactive material is propelled to, and mixed within, the outer regions. Our result supports the theoretical expectation that the radioactive content in the outflow of a type Ib supernova (SN) is thoroughly mixed.  We fit our model to six additional SNe Ibc, of which the majority of the SNe Ib are best described by the two-zone model, and the majority of the SNe Ic by the one-zone model.  Of the SNe Ic, only SN 2007gr was best fit by the two-zone model, indicating that the lack of helium spectral features for this event cannot be attributed to poor mixing.

\end{abstract}


\begin{figure*}
 \centering
 \includegraphics[bb=0 0 256 186, scale=1.4]{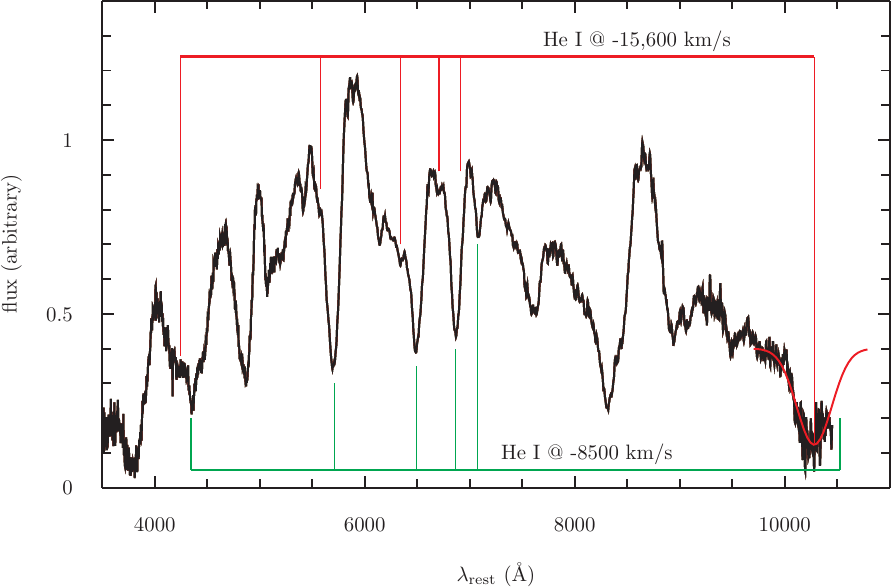}
 \caption{Rest-frame spectrum of SN 1999dn (Ib) at $+17$ days (17-September 1999), obtained by Matheson et al. (2001).  A single Gaussian (in red) has been fit to the spectra in the range 9800$\rightarrow$10,500 \AA.  The best-fitting parameters of the fit give a central wavelength of 10,281 \AA $ $, which if this line is due to He I $\lambda$10,830, corresponds to a velocity of $v_{\rm He} =$ $-15$,576 km/s.  The green lines show He I lines $\lambda$4471, $\lambda$5876, $\lambda$6678, $\lambda$7065 and $\lambda$7281 moving at $-8500$ km/s, while the red lines are the same He I features moving at $-15$,576 km/s.  Two absorption features are seen at $\sim$6400 \AA $ $  and $\sim$6700 \AA, which may be due to blueshifted He I $\lambda$6678 and $\lambda$7065 at the (higher) velocity determined from the Gaussian fit.  A weak feature blueward of the He I $\lambda$5876 feature at $-8500$ km/s is also seen, which is likely due to the same He I line but at $-15$,600 km/s, and it is also seen that the blueshifted absorption line of He I $\lambda$7281 at $-15$,600 km/s is likely blended (on the red side) with the He I $\lambda$7065 line at $-8500$ km/s.  }
\label{fig:gauss_17days}

\end{figure*}

\begin{figure*}
 \centering
 \includegraphics[bb=0 0 256 186, scale=1.4]{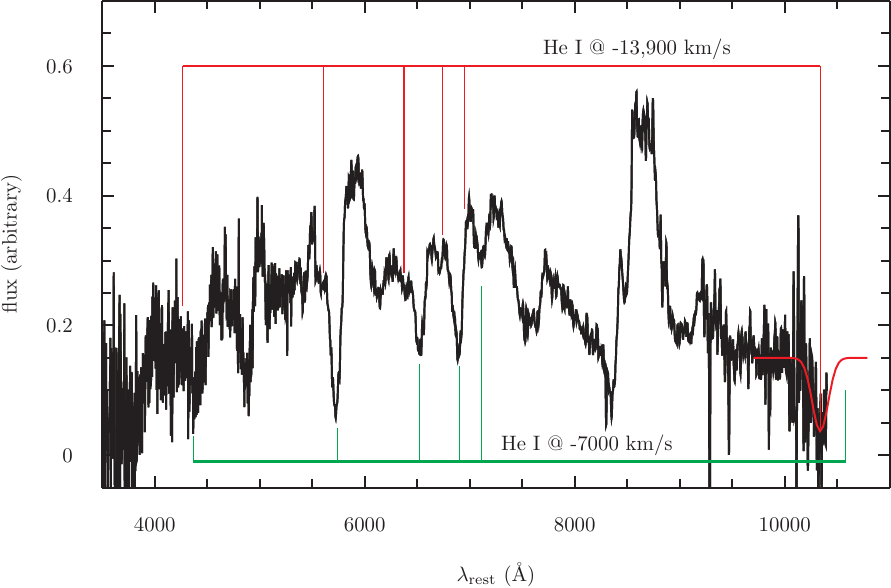}
 \caption{Rest-frame spectrum of SN 1999dn (Ib) at +38 days (08-October 1999), obtained by Matheson et al. (2001).  A single Gaussian (in red) has been fit to the spectra in the range 9,800$\rightarrow$10,500 \AA.  The best-fitting parameters of the fit give a central wavelength of 10,339 \AA, which if this line is due to He I $\lambda$10,830, corresponds to a velocity of $v_{\rm He} =$ $-13$,909 km/s.  The solid green lines are the same He I lines as in Fig. \ref{fig:gauss_17days}, but this time moving at $-7000$ km/s, while the red lines are the same He I features moving at $-13$,909 km/s.  The ejecta velocities have clearly decreased between the two epochs.  The same two absorption features at $\sim$6400 \AA $ $  and $\sim$6700 \AA $ $ that were seen in the $+17$ days spectrum are still there, and their velocity is consistent with that determined from the Gaussian fit, although the feature at $\sim$6700 \AA $ $ is better fit by a slightly higher velocity of $\sim$ $-15$,500 km/s. An additional absorption feature is seen at $\sim$5600 \AA $ $, which may be due to blueshifted He I $\lambda$5876. }
\label{fig:gauss_38days}

\end{figure*}

\section{Introduction}

Type Ib and Ic supernovae (SNe) arise from massive stars whose outer envelopes of hydrogen (Ib) and helium (Ic) have been stripped, either partially or completely, before exploding.  SNe Ibc are observationally classified by their optical spectra (e.g. Filippenko 1997), where SNe Ib exhibit strong He I lines, while SNe Ic exhibit very weak or no He I lines.  The spectra of SNe Ib and Ic have shown that both subclasses are quite heterogeneous, with varying strengths of the He I lines in SNe Ib, and the occasional presence of weak He I lines in the latter.  It has thus been suggested by Filippenko (1997) that rather than being distinctly different subclasses, it may be more appropriate to consider that a continuum of helium strengths exist among SNe Ibc, and that the presence or lack of He I lines in the spectra simply reflects the physical conditions within the SN.

Studies have shown that helium lines arise via non-thermal excitation (e.g. Li \& McCray 1995; Li et al. 2012), and require a departure from local thermodynamic equilibrium (LTE; e.g. Harkness et al. 1987; Lucy 1991; Swartz et al. 1993; Dessart et al. 2011).  High-energy $\gamma$-rays are produced during the radioactive decay of nickel into cobalt and then into iron ($^{56}_{28}$Ni $\rightarrow$ $^{56}_{27}$Co $\rightarrow$ $^{56}_{26}$Fe), which Compton scatter with free and bound electrons, ultimately producing high-energy electrons that deposit their energy in the ejecta through heating, excitation and ionization.  Building upon previous work done by Woosley et al. (1995), recent theoretical models (e.g. Dessart et al. 2011, 2012; Li et al. 2012) have found that while SNe Ic display very weak/no He I lines, this may not be due simply to helium deficiency in SNe Ic, but rather the requirement for producing He I lines is that the nickel is mixed thoroughly enough so that the helium and nickel are within a $\gamma$-ray mean free path.

Two schools of thought currently exist regarding the lack of strong helium spectral lines for SNe Ic: either the helium is present but essentially hidden because it is not excited due to poor mixing (Woosley et al. 1995; Dessart et al. 2011, 2012; Li et al. 2012), or simply the helium is not present.  However, Hachinger et al. (2012) find that only a small amount of helium needs to be present ($0.06-0.14$ $\rm M_{\odot}$) to then be observed, implying that it is not possible to hide very much helium in the ejecta.  Observations are clearly needed to try and distinguish between these theoretical scenarios.

In fact He I lines have been observed in some SNe Ic.  Type Ic SN 1997B displayed a strong absorption feature in the near-infrared (NIR) that was attributed by Clocchiatti et al. (2004) as due to the He I $\lambda$10,830 triplet.  Filippenko et al. (1995) found strong, blueshifted He I $\lambda$10,830 during the first month post maximum of SN 1994I.  Clocchiatti et al. (1996) presented evidence showing that the centroid of the sodium doublet ($\lambda$5890, $\lambda$5896; hereafter Na I D) in 1987M and 1994I shifts to the blue a few weeks after maximum, which is likely due to a growing He I $\lambda$5876 line blending on the blue side of Na I D absorption feature.  However it was also argued by Wheeler et al. (1994) that the spectra of 1987M and 1994I are also consistent with the complete absence of He I, with much less than 0.1 $\rm M_{\odot}$ of hydrogen and helium in the ejecta.  Finally, Matheson et al. (2001) found no compelling evidence for the presence of helium lines in a large sample of SNe Ic 
spectra.

Interpreting the absorption feature redward of 10,000 \AA $ $ in Ibc spectra has proved to be controversial, and it is not completely accepted whether this line is solely due to the He I $\lambda$10,830 triplet, whether it is blended with other atoms/ions, or whether it is entirely due to alternative ions.  Millard et al. (1999) fit the $\lambda$10,250 absorption feature of SN 1994I with a combination of C I $\lambda$10,695 and He I $\lambda$10,830, with the latter detached at 18,000 km/s.  Gerardy et al. (2002) found that the absorption feature around 10,500 \AA $ $ for SN 2002ap (Ic-BL; broad-lined Ic) and SN 2000we (Ic) was best fit by a blend of Mg II, C II and Si I.  The same authors did find that the He I triplet was a good fit to the 10,500 \AA $ $ feature in SN 2001B (Ib).  Moreover, Mazzali \& Lucy (1998) have also suggested that the $\lambda$10,830 absorption feature can be Mg II $\lambda$10,910, though this was for type Ia SN 1994D. It therefore appears that this near-infrared (NIR) absorption 
feature can be due to many atoms and/or ions whose relative contribution varies from event to event.

In this paper we have attempted to determine the origins of this absorption feature in a type Ib supernova (SN 1999dn) in a very straight-forward manner, and in doing so we have also determined the distribution of the radioactive elements within the ejecta.  We make the assumption that the absorption feature redward of 10,000 \AA $ $ is He I $\lambda$10,830, and then calculate its blueshifted velocity.  We then look for other He I absorption features at this velocity.  If He I absorption features in the visible (VIS) part of the spectra are seen at this blueshifted velocity, then we regard this as strong evidence that the NIR absorption feature is He I.  Here we have inspected two epochs of spectra taken of type Ib SN 1999dn by Matheson et al. (2001), which occurred in nearby galaxy NGC 7714.  Falco et al. (1999) measured the redshift of NGC 7714, finding $z=0.00935$, which is the value used in this paper.

In Section \ref{section:spectra} we search for blueshifted He I lines in both spectra, and then model the first epoch using SYN++ (Thomas et al. 2011).  In Section \ref{section:bolometric} we model the bolometric light curve (LC) of SN 1999dn constructed from the photometry obtained by Benetti et al. (2011) using a one- and two-zone model originally presented in Maeda et al. (2003), to determine the distribution of nickel within the ejecta. We model six additional SNe Ibc and then discuss our results in Section \ref{section:discussion}, and summarize our conclusions in Section \ref{section:conclusions}.

\section{Spectra}
\label{section:spectra}

\begin{figure*}
 \centering
 \includegraphics[bb=0 0 256 186, scale=1.4]{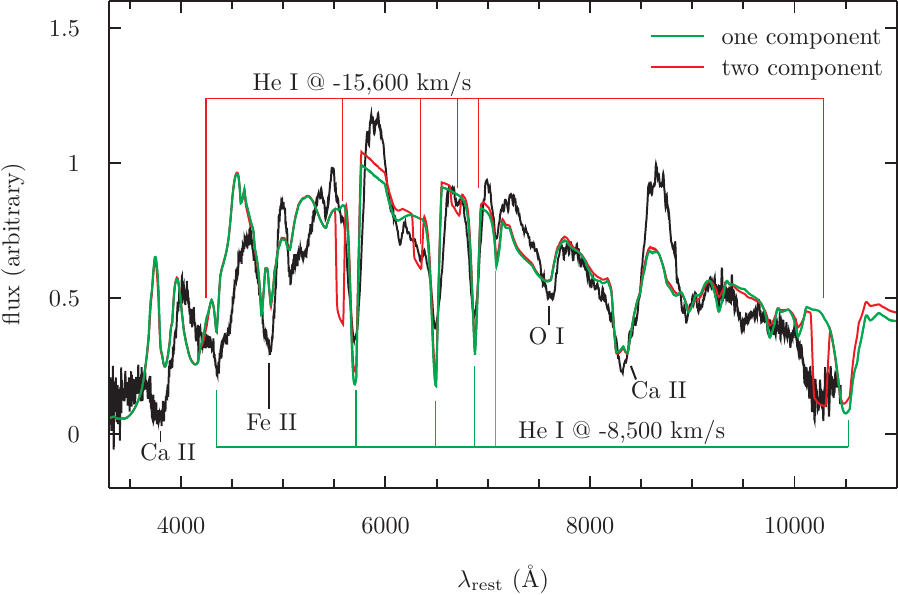}
 \caption{Rest-frame spectrum of SN 1999dn (Ib) at +17 days obtained by Matheson et al. (2001), which has been modelled with synthetic spectra created with SYN++.  The blackbody temperature is $T_{\rm bb} = $  5000 K and the photospheric velocity is $v_{\rm phot} = $ 6000 km/s.  Two synthetic spectra are shown, one with only a single He I component at a velocity of -8500 km/s (green), and another with two velocity components at $-8500$ km/s and $-15$,600 km/s (red).  All other atoms/ions (Fe II, O I, Ca II, Ti II) are at the photospheric velocity.  The two-component synthetic spectrum reproduces the weak absorption lines around 6400 and 6700 \AA$ $, which are due to blueshifted He I $\lambda$6678 and $\lambda$7065 at $-15$,600 km/s.  These lines are not present in the single-component model.  Note that the two-component model over-predicts the absorption line at blueshifted He I $\lambda$5876, which is a combination of He I and Fe II.  The opacity of the higher-velocity He I component in the two-component model is lower than the slower component, which is the same as in the bolometric modelling, where the inner, slower component also has a higher opacity than the outer, quicker component.  All atoms/ions are labelled in the figure, apart from Ti II, which accounts for line-blanketing at wavelengths blueward of 6000 \AA.}
\label{fig:spectra_SYN}

\end{figure*}

Spectra of SN 1999dn, originally presented in Matheson et al. (2001), were downloaded from the WiseREP SN spectrum database (Yaron \& Gal-Yam, 2012)\footnote{http://www.weizmann.ac.il/astrophysics/wiserep/}. The two epochs investigated here both display the NIR absorption feature of interest, and are for phases $+17$ days (17-September 1999) and +38 days (08-October 1999) from R-band maximum brightness, using the terminology from Matheson et al. (2001).  The spectra have been transformed to rest-frame, but they have not been corrected for rest-frame extinction.

For both epochs a single Gaussian was fit to this absorption feature using a program written in Pyxplot\footnote{http://pyxplot.org.uk}.  While neither spectrum extends far enough into the infrared to show the \textit{entire} absorption feature, in both epochs a clear trough is observed as well as a slight increase in flux at redder wavelengths.  This behaviour is more clearly pronounced in the $+17$ days spectrum, but is still evident in the later epoch.

\subsection{+17 days}

A single Gaussian (in red) has been fit to the spectra in the range 9800$\rightarrow$10,500 \AA.  The best-fitting parameters of the fit give a central wavelength of $\lambda_{\rm He}=$10,281 \AA $ $, which if this line is due to He I $\lambda$10,830, corresponds to a blueshifted velocity of $v_{\rm He} = $ $-15$,576 km/s.  The solid green lines show He I lines $\lambda$4471, $\lambda$5876, $\lambda$6678, $\lambda$7065 and $\lambda$7281 moving at $-8500$ km/s, while the dashed red lines are the same He I features moving at $-15,600$ km/s.

In the $+17$ days spectrum clear absorption features are seen at $\sim$6,400 \AA $ $ and $\sim$6,700 \AA, which are likely due to blueshifted He I $\lambda$6678 and $\lambda$7065 at the (larger) velocity determined from the Gaussian fit.  A weak feature blueward of the He I $\lambda$5876 feature at $-8500$ km/s is also seen, which is likely due to the same He I line but at $-15$,600 km/s.

While a clear absorption feature is seen for He I $\lambda$4471 at $-8500$ km/s, no corresponding absorption feature is seen at $-15$,600 km/s, which is likely due to it being blended with strong Fe II and Ti II lines.   It is also seen that the blueshifted absorption line of He I $\lambda$7281 at $-15$,600 km/s is likely blended (on the red side) with the He I $\lambda$7065 line at $-8500$ km/s.

\subsection{+38 days}

At $+38$ days the same weak He I absorption features are seen at the higher velocity determined from the Gaussian fit.  This time the central wavelength is found to be $\lambda_{\rm He}=$10,339 \AA, which if this line is due to He I $\lambda$10,830, corresponds to a blueshifted velocity of $v_{\rm He} =$ $-13$,909 km/s.  The central wavelength has moved to the redder part of the spectrum, indicating that the velocity has decreased between the two epochs.

The same two absorption features at $\sim$6400 \AA $ $  and $\sim$6700 \AA $ $ that were seen in the $+17$ days spectrum are still there, and their velocity is consistent with that determined from the Gaussian fit, although the feature at $\sim$6700 \AA $ $ is better fit by a slightly higher velocity of $\sim$ $-15$,500 km/s.  Another absorption feature is seen at $\sim$5600 \AA $ $, which may be due to blueshifted He I $\lambda$5876 at $-13$,909 km/s.

In conclusion, in both epochs two helium components are seen in the spectra: one corresponding to the stronger, and slower moving He I lines, and another higher velocity component that is consistent with the velocity determined from the Gaussian fit of the absorption feature redward of 10,000 \AA $ $.

\subsection{Spectral modelling}

The investigation presented in the previous sections has shown that there is a strong possibility that there are two velocity components for He I in the spectra of SN 1999dn.  Next, we modelled the spectrum from $+17$ days with SYN++ to see if we can reproduce the He I features at the two velocities.  As mentioned in the previous section, we have not corrected for rest-frame extinction, which has a small value (E(B-V)$_{\rm rest}$=0.048 mag) and will not affect the SYN++ fit significantly.

Plotted in Fig. \ref{fig:spectra_SYN} are two synthetic spectra, one with a single He I component fit at a blueshifted velocity of $-8500$ km/s (green), which is detached from the photospheric velocity of $v_{\rm phot} = $ 6000 km/s, and the other with an additional velocity component, also detached, at $-15$,600 km/s (red).  In the SYN++ modelling, the velocity of the higher velocity component is restricted to the range 15,600 to 20,000 km/s.  The photospheric velocity and blackbody temperature ($T_{\rm bb} = $  5000 K) are identical in both model spectra.  With regards to the modelling parameters, both spectra are identical apart from the inclusion of a second He I component in the red spectrum.  The rest of the lines in the spectra are due to Fe II, O I, Ca II, Ti II and Si II.  The Sobolev opacity of the He I line at the slower(faster) velocities are log($\tau$)=0.4($-0.4$), which is computed using a strong reference line (for He I is $\lambda$7771.94 \AA) that the other He I lines are normalized against.  In this simple model the opacity of the slower component is higher than that of the more rapid component.

It is seen that the spectrum that includes a second, higher-velocity He I component reproduces the weak absorption lines around 6400 and 6700 \AA$ $, which are due to blueshifed He I $\lambda$6678 and $\lambda$7065 at $-15$,600 km/s, and are not present in the single-component model.  It is seen that the synthetic blueshifted He I $\lambda$5876 absorption feature in the two-component model is much stronger than the actual spectrum, and is probably a blend of blueshifted He I at $-15$,600 km/s and Fe II at the photospheric velocity.  We also note that SYN++ assumes LTE level populations whereas He I lines are non-thermally excited.  It is therefore not unexpected that the He I fluxes are not correctly reproduced.  In particular the 2s3s stare is metastable and hence $\lambda$10,830 will have a much larger optical depth than, for example, $\lambda$5876.  Finally, no absorption feature is seen for blueshifted He I $\lambda$4471 in the two-component synthetic spectrum which agrees with the results from section \ref{section:spectra}.  

Branch et al. (2002) performed an in-depth analysis of the +17 days spectrum of SN 1999dn.  Using SYNOW they found a good fit to the spectrum using $v_{\rm phot} = $ 6000 km/s and $T_{\rm bb} = $ 4800 K, and the same elements listed above.  In their spectrum, He I was detached at $-8000$ km/s, which is a close match to the velocity determined here for the slower He I component.  Encouragingly they also predict a blueshifted He I $\lambda$10,830 line at $-8000$ km/s with their SYNOW model (see Fig. 3 in Branch et al. 2002). 

Additionally, Deng et al. (2000) modelled low-resolution spectra of SN 1999dn at three epochs, the last one being 14-September 1999 (i.e. three days before the epoch modelled here).  Using SYNOW, they found $v_{\rm phot} = $ 9000 km/s and $T_{\rm bb} = $ 5300 K, which are consistent with the parameters derived here and by Branch et al. (2002).  However, unlike our study, Deng et al. (2000) have included many more atoms and ions in their modelling: in addition to those listed above, they also used Mg II, [O II], C I, CII, Na I, Ca I, Ni II and H I.  

Interestingly, Deng et al. (2000) have also questioned whether two different C II velocity components exist in the helium layer, with the ions detached at $-20$,000 km/s at early times, and at roughly $-10$,000 km/s on the 14-September.  However, the authors favour a different scenario in which the absorption feature near 6300 \AA $ $ is due to H$\alpha$ in the early epochs, at $\sim -19$,000 km/s, while in the 14-September epoch the same feature is C II (6580 \AA), thus concluding that a thin high-velocity hydrogen layer exits outside of the slower helium layer in SN 1999dn.

The results of the SYN++ modelling generally reproduce the results seen in the preceding section, namely that it is likely that there are two components of He I in the outflow of SN 1999dn.  The clearest evidence arises from the blueshifted He I features at $-15$,600 km/s, especially blueshifted He I $\lambda$6678 and $\lambda$7065, which are only present in the two-component model.

\section{Bolometrics}
\label{section:bolometric}

\subsection{Time of Explosion}

By comparing the rise time of SN 1999dn with those of other SNe Ibc, Benetti et al. (2011) estimated the date of explosion to have occurred $\approx$ 16 August 1999 (JD=2451406.0).

Using equation 17 of Piro \& Nakar (2013), it is possible to put a lower limit on the explosion date by using a measurement of the photospheric velocity and temperature at a single epoch during the LC rise.  Benetti et al. (2011) modelled several epochs of spectroscopy obtained of 1999dn using SYNOW, with two epochs occurring before maximum light ($-6.0$ and $-2.3$ days).  Using their results from $-2.3$ days (JD=2451415.7): $T_{\rm bb}\sim 9100$ K, $v_{\rm ph}\sim$10,400 km/s, and the luminosity determined from our bolometric LC ($L$ $\sim 1.05 \times 10^{42}$ erg/s), we find $t_{\rm min}=5.1$ days.  This implies a lower limit to the explosion date of JD=2451410.6, or $\approx$ 20 August 1999.

We have also used the results from the spectrum modelled by Benetti et al. (2011) for the epoch $-6$ days (JD=2451413.5): $T_{\rm bb}\sim 7000$ K, $v_{\rm ph}\sim$15,000 km/s.  We have estimated the luminosity at this epoch (which extend further back in time than our actual bolometric LC) by fitting a $t^{2}$ curve (Arnett 1982) to the early bolometric LC, and then extracting the luminosity at JD=2451413.5, to which we find $L$ $\sim 8.4 \times 10^{41}$ erg/s.  This implies $t_{\rm min}=5.3$ days before this measurement, and a lower limit to the explosion date of JD=2451408.2, which is $\approx$ 17 August 1999. This epoch predicts an earlier limit to the date of explosion compared with the latter epoch, and within two days of the explosion date estimated by Benetti et al. (2011).  Taking into consideration the results of Benetti et al. (2011) and those found here, throughout this paper we have used an explosion date of JD=2451408.0$\pm$2.0 days.

\subsection{Construction of the bolometric light curve}
\label{section:bolo_construct}

We collected the photometry published by Benetti et al. (2011), and using their value of the total extinction along the line of site (E(B-V)$_{\rm total}=0.10 \pm 0.05$ mag) created a quasi-bolometric LC in filters $UBVRI$.  While the bulk of radiation emitted by a given SN is emitted primarily in the optical bands, work by e.g. Tomita et al. (2006), Modjaz et al. (2009) and Cano et al. (2011) have shown that roughly 50--70$\%$ of the radiation is emitted within $UBVRI$, with a sizable contribution early on in the UV, and an increasing contribution in the IR well after the peak.  Therefore our derived bolometric LC may have a large systematic uncertainty of order $\sim20\%$.  We note that due to the paucity of IR observations of SN 1999dn, as well as a desire to keep the analysis as simple as possible without introducing additional sources of uncertainty, we have not attempted to estimate the IR contribution as we have done for other work (e.g. Cano et al. 2011).

Our general procedure for creating the bolometric LC (as well as those presented in Section \ref{section:model_Ibc}) is:

\begin{enumerate}
 \item Collect published $UBVRI$ photometry.
 \item Correct for extinction along the entire line of sight (foreground and rest-frame).
 \item Convert magnitudes into monochromatic fluxes using flux zeropoints in Fukugita (1995).
 \item For each epoch of multi-band observations (i.e. for each spectral energy distribution, SED), using the effective wavelengths from Fukugita (1995):
 \subitem   i. Interpolate (linearly) between each datapoint.
 \subitem  ii. Integrate the SED over frequency, assuming zero flux at the integration limits.
 \subitem iii. Correct for ``filter overlap''.
\end{enumerate}

The linear interpolation and integration were performed using a program written in Pyxplot.  In order to make the most of the published photometry, when there was an epoch without contemporaneous $UBVRI$ photometry, we interpolated the individual LCs (using a linear interpolation) to estimate the amount of ``missing'' flux in a given filter.  The resulting fluxes in units (mJy Hz) are converted into units of (erg/s/m$^{2}$), and then into luminosities using the distances derived from the observed redshifts which were calculated using the latest cosmological parameters determined by Planck (Planck Collaboration et al., 2013; $H_{0} = 67.3$ km s$^{-1}$ Mpc$^{-1}$, $\Omega_{M} = 0.315$, $\Omega_{\Lambda} = 0.685$).  

Peak bolometric light was determined using a program written in Python that fits multi-order polynomials to the bolometric LC.  The best-fit parameters are found using scipy.optimize.leastsq, which minimizes the sum of squares to the fitted polynomial.  The peak time is then found using scipy.optimize.fmin to find the minimum of the function, which in this case is the negative of the fitted polynomial.  The program was the run several times over different time ranges (i.e. excluding arbitrarily chosen datapoints) to determine the statistical uncertainty of the derived peak time.  The total error in the peak time is therefore a combination of the uncertainty in the explosion date as well as the statistical uncertainty from the fit, which we have added in quadrature.

Taking the date of explosion to be JD=2451408.0$\pm$2.0 days, we find a peak time of $t_{\rm peak} = 10.14 \pm 2.04$ days.  The luminosity at peak light is found to be $L_{\rm peak}=1.13 \times 10^{42}$ erg/s. 

\begin{figure*}
 \centering
 \includegraphics[bb=0 0 568 190, scale=0.89]{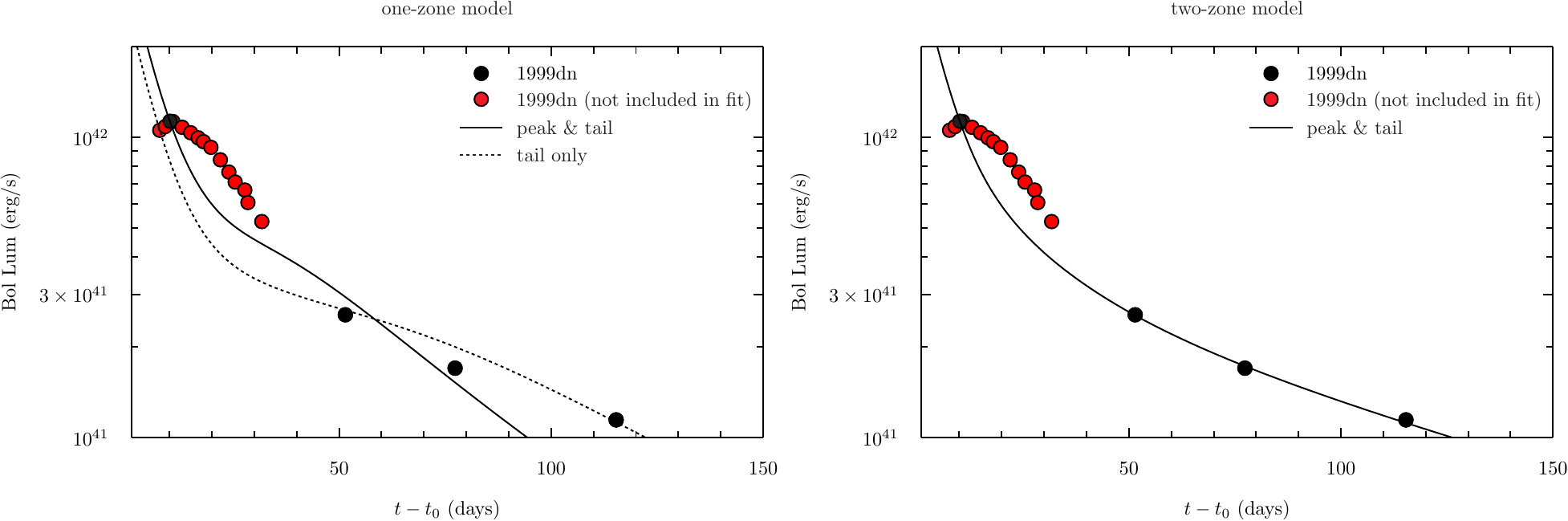}
 \caption{Bolometric light curve of SN 1999dn.  \textit{Left:} The peak+tail (solid) and tail (dotted) have been fit with the one-zone model from Maeda et al. (2003) for all data at times $>$ 50 days.  For the one-zone model to be deemed a good fit to the bolometric LC, the nickel mass found from fitting the peak+tail must be approximately the same as that from fitting only the tail.  It is found that the nickel mass and opacity found from fitting just the tail do not provide a good fit to the peak+tail data ($R^{2}=0.748$), which is also reflected by the fact that the dotted line does not dissect the peak of the LC.  It is also seen that the solid line in the one-zone model does not reproduce the tail very well.  Instead the two-zone model (\textit{right}) provides a much better fit to the peak+tail, both visually and statistically ($R^{2}=0.998$).  The nickel masses and opacities found from fitting both models are displayed in Table \ref{table:params}.   }
\label{fig:1999dn}

\end{figure*}

\subsection{One-zone model}
\label{section:one-zone}

Using an adaptation of the Arnett (1982) model, which includes the energy produced by the decay of nickel into cobalt, and then into iron, we have estimated the mass of nickel nucleosynthesized during the explosion to be $M_{\rm Ni}=0.040 \pm 0.005$ $\rm M_{\odot}$.  The uncertainty in the nickel mass is statistical only, and is dominated primarily by the uncertainty in the explosion date.  The median nickel mass determined by Cano (2013) for a SN Ib is $M_{\rm Ni}=0.16$ $\rm M_{\odot}$ (though over a larger frequency interval, and includes IR photometry), thus it appears SN 1999dn synthesized much less nickel than the ``typical'' SN Ib.

As the bolometric LC extends to over 115 days, we attempted to constrain the nickel mass from the exponential tail using a simple one-zone model (Maeda et al. 2003; M03 hereon).  M03 investigated the issue of $\gamma$-ray transport in SN ejecta using a simple model, building upon work done previously by Clocchiatti \& Wheeler (1997).  The decay of nickel into cobalt, and then into iron produces $\gamma$-rays and positrons, of which a fraction of the former are thermalized and deposit their energy into the SN ejecta.  As with the Arnett model, we have included the energy deposited by the decay of nickel and cobalt into their daughter products:

\begin{equation}
\label{equ:single-model_co}
L(t) =  M_{\rm Ni}(e^{-t/t_{\rm Co}}[\epsilon_{\rm Co}(1-e^{-\tau}) + \epsilon_{e^{+}}])
\end{equation}

and 
\begin{equation}
\label{equ:single-model_ni}
L(t) =  M_{\rm Ni}(e^{-t/t_{\rm Ni}}[\epsilon_{\rm Ni}(1-e^{-\tau})])
\end{equation}

\noindent where time $t$ is expressed in units of days, $t_{\rm Co}$ is the e-folding time of $^{56}$Co (113 days; M03) and $M_{\rm Ni}$ is the mass of nickel produced during the explosion in units of grams.  $\tau=\kappa_{\gamma}\rho R = \tau_{0} / t^{2}$ is the optical depth to $\gamma$-rays, where $\tau_{0}$ is the optical depth to $\gamma$-rays at $t=1$ day.  The energy input from $\gamma$-rays and positrons are $\epsilon_{\gamma} = 6.8 \times 10^{9}$ erg/s/g and $\epsilon_{e^{+}} = 2.4 \times 10^{8}$ erg/s/g respectively.  Positron channels are negligible for the decay of nickel into cobalt, so this term doesn't appear in equation \ref{equ:single-model_ni}.  We have also assumed that the same optical depth applies to both decay reactions, though in reality the optical depths for $^{56}_{28}$Ni $\rightarrow$ $^{56}_{27}$Co, and  $^{56}_{27}$Co $\rightarrow$ $^{56}_{26}$Fe differ due to the different line energies of the emitted $\gamma$-rays; however, given the large optical depth early on this effect should be negligible.  By adding equations \ref{equ:single-model_co} and \ref{equ:single-model_ni} and then fitting them to the bolometric LC it is possible to determine the amount of nickel nucleosynthesized during the explosion, where $M_{\rm Ni}$ and $\tau_{0}$ are the free-parameters during the fit.

When performing the fit we are interested in obtaining a good fit to the bolometric peak as well as the late-time tail.  The phase between the peak and the tail, which we will refer to as the ``slope'', is less important when fitting the data because during this period energy that was trapped before the peak is still being released as the photosphere recedes into the ejecta, and the SN expands, decreasing the overall optical depth.  Therefore it is expected that the luminosity predicted by our single-zone model is less than the observations.  It should be considered however that this model should be able to successfully recover the peak luminosity from fitting of only the tail phase.

A key uncertainty then is determining when the ``slope'' phase transitions into the ``tail'' phase.  As such we have fit the bolometric LC of SN 1999dn at $t-t_{0} > 50$ and $60$ days with a Python program that uses scipy.optimize.leastsq to find the best-fit parameters for each model.  Our fitting procedure is to fit: (1) the peak\footnote{This is the fiducial peak determined from our python program.}+tail simultaneously to determine $M_{\rm Ni}$ and $\tau_{0}$, (2) just the tail to determine $M_{\rm Ni}$ and $\tau_{0}$, and then compare.

For the one-zone model to be considered a good fit to the bolometric LC, the nickel masses found from fitting just the tail must be similar to that found from fitting the peak+tail.  This is because even though the fit obtained from fitting the peak+tail may appear to be a good statistical fit to the data, the tail can be poorly reproduced (e.g. Fig. \ref{fig:1999dn} on the left).  From a theoretical standpoint it is important to obtain a very good fit to the slope in order to reproduce the exact amount of nickel created during the explosion.

For each fit we have estimated the goodness of fit by calculating the correlation coefficient, $R^{2}$, for: (1) the peak+tail simultaneously, (2) just the tail, and (3) the peak+tail using the fitted values determined from fitting just the tail.  Thus if the values obtained from fitting just the tail are a good match to those from fitting the peak+tail, the correlation coefficient will reflect this as a good statistical fit.  We have estimated the errors in the best-fit parameters by performing the fit several times to account for the uncertainty in the date of explosion.  As such we have modelled three bolometric LCs of SN 1999dn, each with a different date of explosion (i.e. the minimum, maximum and exact date of explosion).  We then calculated the average value of each parameter, and used the spread of values as an indication of the statistical error derived from the fit.  Our results are listed in Table \ref{table:params}, and the models are displayed in Fig. \ref{fig:1999dn}.

What becomes quickly apparent is that the nickel mass obtained from fitting just the tail ($M_{\rm Ni} = 0.024-0.030$ $\rm M_{\odot}$) is less than that obtained from fitting the peak+tail simultaneously ($M_{\rm Ni} = 0.041$ $\rm M_{\odot}$), and for which we recover the nickel mass obtained from the Arnett model.  It is also seen that the value of $R^{2}$ calculated from fitting the peak+tail when using the nickel mass and optical depth obtained from fitting just the tail are much lower than those determined from the other fits, while inspection of Fig. \ref{fig:1999dn} (left) shows the poor visual fit of the tail when the one-zone model is fit to the tail+peak simultaneously.  This behaviour is consistently found over both time-ranges.  This leads us to conclude that the single-zone model provides a poor physical description of the configuration of the ejecta.

\subsection{Two-zone model}
\label{section:two-zone}
The single-zone model of M03 is an approximation that considers a single homogeneous sphere with a constant density.  However, in reality the geometry of many SNe may not be perfectly symmetric.  One solution can be to consider a homogeneous bi-polar cone with an opening angle $\theta$.  Another solution is the one proposed by M03: a simple one-dimension (1D) model that approximates the configuration of the ejecta to be in two regions.  In a 1D explosion most of the material is expelled at a high velocity, leaving a central, low-velocity region at quite low density.  This is seen especially for cases where the kinetic energy is large, such as those measured for hypernovae ($E_{\rm ke} > 10^{52}$ erg).  As such, 1D models are only able to account for the total inflow/outflow in a single direction.  2D and 3D explosions are not limited by this however, where it is possible to have a inflow and outflow simultaneously with different velocities.  The consequence of this is a central region with a higher density/optical depth than is possible in simple 1D explosion models, and an outer region of smaller density/optical depth.  Thus the two-zone model provides a measure of the amount of mixing occurring in the ejecta, as well as including a degree of the asymmetry in the explosion, both of which are consistent with hydrodynamic models of strongly jetted explosions (e.g. Maeda \& Nomoto 2003).

The energy deposited in the ejecta by $\gamma$-rays and positrons from the decay of cobalt into iron then becomes:

\begin{multline}
\label{equ:double-model}
L(t) =  M_{\rm Ni, in}e^{-t/t_{\rm Co}}[\epsilon_{Co}(1-e^{-\tau_{\rm in}}) + \epsilon_{e^{+}}]\ + \\
M_{\rm Ni, out}e^{-t/t_{\rm Co}}[\epsilon_{Co}(1-e^{-\tau_{\rm out}}) + \epsilon_{e^{+}}]
\end{multline}

\noindent where the variables have the same meaning as in equation \ref{equ:single-model_co}, but this time we consider two deposits of $^{56}$Ni (an inner and an outer).  The energy deposited by the decay of nickel into cobalt in the two zones can be similarly approximated as in Section \ref{section:one-zone} but without the positron contribution.  Both contributions are then added together and then fit to the bolometric LC, for which there are now four free parameters: $M_{\rm Ni, in}$, $M_{\rm Ni, out}$, $\tau_{\rm 0,in}$, and $\tau_{\rm 0,out}$.

We fit the two-zone model to the bolometric LC of SN 1999dn to $t-t_{0} > 50$ days, but due to more free-parameters than datapoints we were not able to fit for $t-t_{0} > 60$ days.  Here we find that the total nickel mass ejected, $M_{\rm Ni, total} = M_{\rm Ni, in} + M_{\rm Ni, out} = 0.040 \pm 0.008$ $\rm M_{\odot}$, which agrees with that determined from the one-zone model (for peak+tail) and the Arnett model.  It is also seen that the opacity of the inner region ($\tau_{\rm 0,in} = 404$,000 $\pm 80$,000) is much higher than the outer region ($\tau_{\rm 0,out} = 1394 \pm 349$), with values that are similar to seen from fitting the hypernovae in M03.  Here the quoted errors are statistical, and arise predominantly from the uncertainty in the explosion date.

Some exciting conclusions may be drawn from our results: (1) the outflow of SN 1999dn is better fit by the two-zone model, implying that the ejecta is likely asymmetric, and (2) similar amounts of nickel are found in the inner ($M_{\rm Ni, in} = 0.019$ $\rm M_{\odot}$) and outer ($M_{\rm Ni, out} = 0.021$ $\rm M_{\odot}$ regions, implying that the outer regions of the ejecta are thoroughly mixed with radioactive material.  Encouragingly a similar result was observed by Takaki et al. (2013) for Ib SN 2012au who also fit their bolometric LC with the two-zone model of M03, finding nickel masses in the inner and outer regions of 0.14 $\rm M_{\odot}$ and 0.12 $\rm M_{\odot}$ respectively.  

\begin{table*}
\centering
\setlength{\tabcolsep}{10.0pt}
\caption{Photometry references}
\label{table:SNe_photometry}

  \begin{tabular}{ccccccc}
  \hline
SN	&	Type	&	$z$		&	$\rm E(B-V)_{total}$	&	$t_{\rm peak}$ (days)	& $L_{\rm peak}$ (erg/s)	&	Refs.\\
\hline
1999dn	&	Ib	&	0.00938		&	0.100			&	$10.14 \pm 2.04$	& $1.13 \times 10^{42}$		&	(1),(2)	\\
2007Y	&	Ib	&	0.004657	&	0.102			&	$18.35 \pm 1.51$	& $0.61 \times 10^{42}$		&	(3)	\\
2008D	&	Ib	&	0.007		&	0.622			&	$18.00 \pm 0.46^{*}$	& $0.76 \times 10^{42}$		&	(4)	\\
2009jf	&	Ib	&	0.007942	&	0.113			&	$21.44 \pm 1.07$	& $2.25 \times 10^{42}$		&	(5)	\\
2004aw	&	Ic	&	0.0175		&	0.364			&	$10.01 \pm 2.99$	& $2.24 \times 10^{42}$		&	(6)	\\
2007gr	&	Ic	&	0.001728	&	0.090			&	$12.28 \pm 2.50$	& $0.67 \times 10^{42}$		&	(7)	\\
2011bm	&	Ic	&	0.0221		&	0.064			&	$31.26 \pm 1.50$	& $4.12 \times 10^{42}$		&	(8)	\\

\hline

\end{tabular}
\begin{flushleft}
$^{*}$ The exact explosion time is known, therefore the quoted error arises only from the statistical uncertainty from determining the peak.\\
(1) \cite{Matheson01}; (2) \cite{Benetti2011}; (3) \cite{Stritzinger09}; (4) \cite{Malesani2009}; (5) \cite{Valenti2011}; (6) \cite{Taubenberger06}; (7) \cite{Hunter09}; (8) \cite{Valenti2012}
\end{flushleft}

\end{table*}

\begin{figure*}
 \centering
 \includegraphics[bb=0 0 563 615,scale=0.88]{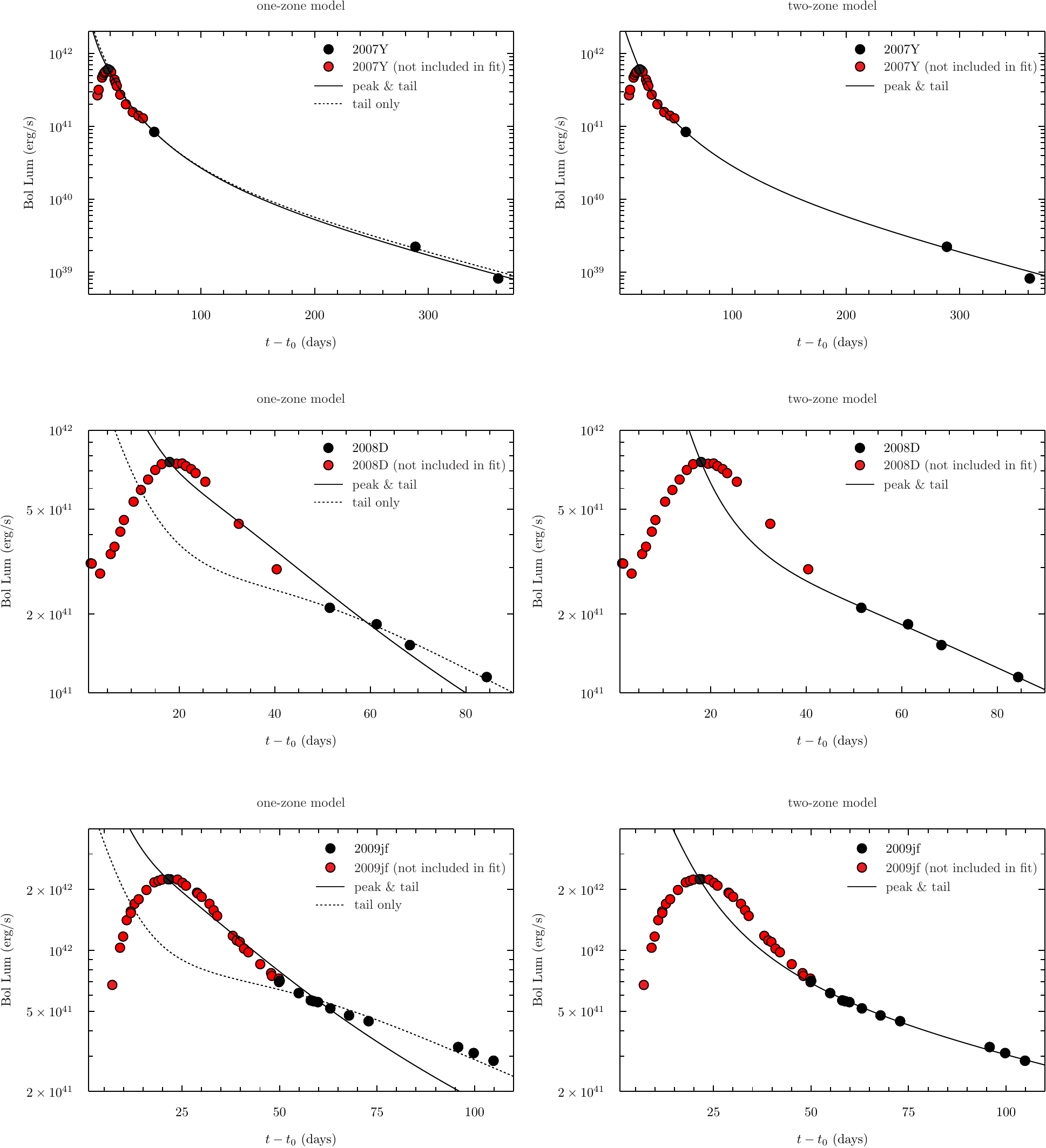}
 \label{figure:Ib}
 \caption{Bolometric LCs of the other SNe Ib in our sample (2007Y, 2008D and 2009jf).  The LCs and models are displayed in the same fashion as Fig. \ref{fig:1999dn}, with the one-zone models on the left and the two-zone models on the right.  The results shown are for times $>$ 50 days.  \textit{Top:} SN 2007Y.  The best-fitting model is uncertain, with the one-zone model providing a good fit for all data $>$ 50 days, but the two-zone model being a marginally better fit for times $>$ 40 and 60 days.  \textit{Middle:} SN 2008D.  The best fitting model is the two-zone model, which estimates that roughly $3-4$ times more nickel is found in the outer layers of the ejecta.  \textit{Bottom:} SN 2009jf.  The best fitting model is the two-zone model, which finds that almost 5 times more nickel is present in the outer regions of the ejecta.  When fitting the one-zone model to the peak+tail for SNe 2008D and 2009jf, visually the tail is poorly fit.  The two-zone model provides a better fit to the bolometric LCs of three of the four SNe Ib considered in this paper, and the best-fitting parameters determined from both models are displayed in Table \ref{table:params}.}
\end{figure*}

\section{Discussion}
\label{section:discussion}

\subsection{Helium lines}

When analyzing the spectra of SN 1999dn, He I absorption lines were seen at two velocities: a slower velocity at which the stronger He I lines were seen, and a larger velocity that was determined by attributing the NIR absorption feature as being due to He I $\lambda$10,830, and fitting a Gaussian to it to determine its blueshifted velocity.  With regards to the on-going debate about the origin of this spectral feature it appears, at least in this event, that it is only He I at a larger blueshifted velocity than the stronger He I lines.  Support for this conclusion comes from the fact that multiple He I lines are seen at the larger velocity in the optical, with the same lines appearing consistently in both epochs.  It is seen that the blueshifted velocity of the $\lambda$10,830 absorption feature decreased between the two epochs, falling from $-15$,576 km/s at 17 past maximum light to $-13$,909 km/s 21 days later. 

Deng et al. (2000) modelled three epochs of spectra of SN 1999dn, with their last epoch being three days younger than the epoch modelled here.  Deng et al. (2000) also concluded that the SN outflow has different velocity components, with a slower helium layer and a more rapid hydrogen layer.  This independent analysis supports the idea that the ejecta is not a single homogeneous blob expanding homologously, but rather is comprised of regions moving with different velocities.

\begin{figure*}
 \centering
 \includegraphics[bb=0 0 563 615,scale=0.88]{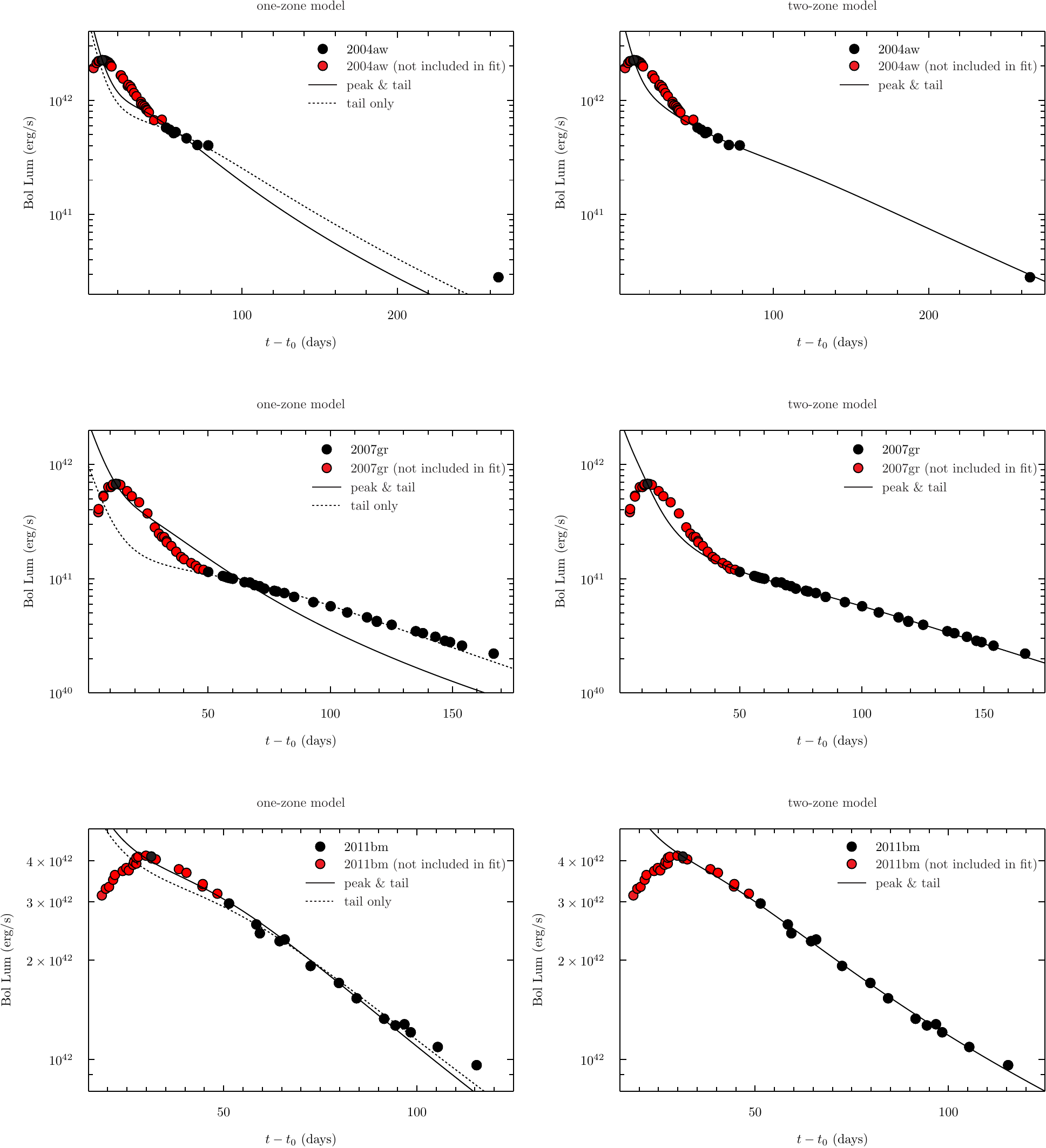}
\label{figure:Ic}
 \caption{Bolometric LCs of the SNe Ic in our sample (2004aw, SN 2007gr and SN 2011bm).  The LCs and models are displayed in the same fashion as Fig.s \ref{fig:1999dn} and \ref{figure:Ib}, with the one-zone models on the left and the two-zone models on the right.  The results shown are for times $>$ 50 days.  \textit{Top:} SN 2004aw.  The best-fitting model is the one-zone model.  \textit{Middle:} SN 2007gr.  The one-zone model (peak+tail) is a poor visual fit to the tail.  Instead, the best fitting model is the two-zone model, which estimates that roughly $2-3$ times more nickel is found in the outer layers of the ejecta.  This result implies that the lack of helium spectral lines cannot be attributed to poor mixing in the ejecta.  \textit{Bottom:} SN 2011bm.  The best fitting model is the one-zone model.  The one-zone model provides a better fit to the bolometric LCs of two of the three SNe Ic considered in this paper, and the best-fitting parameters determined from both models are displayed in Table \ref{table:params}.}
\end{figure*}

\subsection{Two-zone model applied to other SNe Ibc}
\label{section:model_Ibc}

The results from fitting the two-zone model to the bolometric LC of SN 1999dn showed that similar amounts of nickel/cobalt are present in the inner and outer regions of the ejecta.  If we take our results at face value, they imply that the amount of mixing is at least partially due to the ejecta having an asymmetric structure.  Therefore in the case of SN 1999dn, the presence of helium spectral lines can be explained by the ejecta being well mixed and the nickel and helium are located in close proximity in the ejecta.

One wonders however how much can we trust these results when applied to only a single event?  Certainly other authors such as Takaki et al. (2013) have found similar results for another type Ib SN (2012au), but what about other SNe Ibc?  And what about the outstanding uncertainty about the unknown role of mixing in SNe Ib vs. Ic?  Do all SNe Ibc contain similar amounts of helium in the ejecta, but only appear in the spectra of SNe Ib due to mixing brought about via asymmetric explosions?

In an attempt to address these questions we have modelled three additional SNe Ib (SNe 2007Y, 2008D, 2009jf) and three SNe Ic (SNe 2004aw, 2007gr, 2011bm) using the one-and two-zone models.  We chose these particular SNe from the sampled presented in Cano (2013) as both foreground and rest-frame extinction were known, and more crucially, observations of these SNe extended well into the tail phase, therefore making them suitable to be modelled.  The bolometric LCs were constructed using the procedure described in Section \ref{section:bolo_construct}, and we have also determined the time of peak bolometric light for each SNe using our Python program and the explosion dates published in the literature.   Table \ref{table:SNe_photometry} lists the literature references for the photometry, redshifts, total extinction along the line of sight and our derived peak times and luminosities.  We have modelled the SNe using the procedure in Sections \ref{section:one-zone} and \ref{section:two-zone}, where due to the abundance of datapoints for the chosen SNe, we have fit the bolometric LCs over three ranges: $t-t_{0} > $ 40, 50 and 60 days.  As before, the telling statistic to elucidate the best-fitting model is the value of $R^{2}$ from fitting the peak+tail using the nickel mass and opacity determined from fitting only the tail in the single-zone model.  Table \ref{table:params} displays the best-fit values from the two models.

\subsubsection{SN 2007Y (Ib)}

Neither the one- or two-zone modelling consistently proving to be the preferred model for SN 2007Y. For all three time ranges the one-zone model consistently determined that $\approx 0.04-0.05$ $\rm M_{\odot}$ of nickel was synthesized, which is slightly more than that predicted by the Arnett model ($M_{\rm Ni} = 0.034\pm0.003$ $\rm M_{\odot}$).  However,  For times $> 40$ and 60 days the value of the correlation coefficient was quite low when the nickel mass and opacity determined from fitting only the tail in the one-zone model were then applied to the peak+tail in the one-zone model.  At $>$ 40 days the total nickel mass determined from the two-zone model was much greater than that for the single-zone model, though at $>$ 60 days the nickel masses closely agreed.  At $>$ 50 days the one-zone model fit the data very well, with the nickel mass determined from fitting the peak+tail and just the tail in the one-zone model agreeing closely.  

\subsubsection{SN 2008D (Ib)}

The results for SN 2008D are much more conclusive, and indicate that the bolometric LC is best fit by the two-zone model.  The nickel masses derived from the one- and two-zone models agree well ($M_{\rm Ni} \approx 0.047$ $\rm M_{\odot}$), which is slightly more than that determined from the Arnett model ($M_{\rm Ni} = 0.039 \pm 0.001$ $\rm M_{\odot}$).  For all three time-ranges the nickel mass derived from fitting the tail only in the one-zone model was much less than that from fitting the peak+tail, and the value of $R^{2}$ calculated for fitting the peak+tail using the values from fitting just the tail indicated that the the one-zone model was a poor fit, which can also be seen visually.  

The total nickel mass determined from the two-zone model for $>$ 50 days ($M_{\rm Ni} = 0.107 \pm 0.001$ $\rm M_{\odot}$) was much greater than that found in the one-zone model.  For times $>$ 40 and 60 days the total nickel mass was approximately the same as determined from the one-zone model, with both indicating that roughly $3-4$ times more nickel is present in the outer layers compared with the inner.

\subsubsection{SN 2009jf (Ib)}

The bolometric LC of SN 2009jf is best fit by the two-zone model, however the results from all three time ranges are not as consistent as for some of the other SNe.  The nickel masses derived from the one- and two-zone models agree well ($M_{\rm Ni} = \approx 0.163-0.166$ $\rm M_{\odot}$), which is slightly more than that determined from the Arnett model ($M_{\rm Ni} = 0.138 \pm 0.010$ $\rm M_{\odot}$).  For times $>$  40 days the one-zone model could be argued to be a suitable fit to the data, however the one-zone model was a poor fit for the other time ranges.  The total nickel mass derived from the two-zone model for $>$ 60 days is 3 times larger than for any other model, indicating that the one- and two-zone models are both poor fits to the data for times $>$ 60 days.  Moreover, the optical depth of the inner region is very high, $\tau_{0,\rm in} =$($0.1-4$)$\times 10^6$, much higher than is seen for the other SNe for which the two-zone model was the better fit.  This may be indicating that the ejecta of SN 2009jf is complicated and may possess more than two zones.  However, if we take the results of the two-zone model at face value, it then suggests that on average $\approx 5$ times more nickel is present in the outer layers relative to the inner, denser region.

\subsubsection{SN 2004aw (Ic)}

For SN 2004aw the nickel mass and opacity determined from fitting the tail only in the one-zone model, and then applied to the peak+tail provides a good visual and statistical fit for all three time ranges.  Both the one- and two-zone models estimate that $\approx0.08$ $\rm M_{\odot}$ of nickel was created during the explosion, which agrees well with that determined from the Arnett model ($M_{\rm Ni} = 0.077 \pm 0.011$ $\rm M_{\odot}$).  The one-zone model provides a good, consistent fit to the bolometric LC of SN 2004aw.

\subsubsection{SN 2007gr (Ic)}

For all three time ranges the one-zone model was a poor fit, and the bolometric LC of SN 2007gr is best fit by the two-zone model.  Both models predict that $\approx 0.03$ $\rm M_{\odot}$ of nickel was nucleosynthesized during the explosion, which agrees well with the nickel mass derived from the Arnett model ($M_{\rm Ni} = 0.026 \pm 0.003$ $\rm M_{\odot}$).  The value of $R^{2}$ determined by fitting the peak+tail with the values obtained from fitting just the tail in the one-zone model also indicates that the one-zone model is a poor fit, which can also be seen visually.  The two-zone model finds that between $2-3$ times more nickel is found in the outer layers relative to the inner region.

\subsubsection{SN 2011bm (Ic)}

As for SN 2004aw, the bolometric LC of SN 2011bm is well described by the one-zone model.  For all three time ranges, the nickel mass and opacity determined from fitting just the tail in the one-zone model proved to be a good fit for the peak+tail in the one-zone model.  All models estimate that $\approx 0.361-0.369$ $\rm M_{\odot}$ of nickel was synthesized during the explosion, which is only slightly higher than the nickel mass found from the Arnett model ($M_{\rm Ni} = 0.349 \pm 0.022$ $\rm M_{\odot}$).

\subsubsection{General trends}

Some really interesting conclusions can be drawn from the results.  First, from a phenomenological perspective, of the SNe Ib, arguably three out of four are best fit by the two-zone model.  The geometry of SN 2007Y proved more difficult to constrain, with the best-fitting model perhaps being the one-zone model, though this model did not consistently prove to be the best fit to the bolometric LC over all three time ranges.  That the bulk of the SNe Ib are fit by the two-zone model implies that these events have aspherical geometries.  Of the SNe Ic, two are best fit by a one-zone model (SNe 2004aw and 2011bm), while SN 2007gr is better fit by the two-zone model.  In this small sample, two-thirds of the SNe Ic possess spherical symmetry.  For SN 2007gr it appears that the lack of helium spectral features cannot be attributed to poor mixing as the fits over all three time-spans indicate that $2-3$ times more nickel are present in the outer layers of the ejecta.  

From the modelling, as was seen for SN 1999dn, while the value of the correlation coefficient was seen to be quite good for the one-zone model for all of the SNe, visually the tails of SNe 2008D, 2009jf and 2007gr were poorly reproduced.  This is reflected in the much smaller nickel masses estimated from fitting only the tails for these events.  Next, the nickel masses determined from the various models were sometimes more than that found using the Arnett model, which suggests that the ratio of peak bolometric to radioactive luminosities is not unity, but may be slightly higher (e.g. Takaki et al. 2013).  And finally, for all of the SNe whose bolometric LCs were better described by the two-zone model, the inner region always had a higher opacity to $\gamma$-rays than the outer region (see Table \ref{table:params}).


\subsection{Amount of nickel mixing}

The results of the bolometric modelling allow us to comment on the role of mixing in SNe Ibc.  As discussed in the introduction, the lack of He I lines in Ic spectra may either be due to the absence of helium in the ejecta, or that the helium is present but not excited, therefore making it essentially hidden to the observer.  

A very interesting paper by Frey et al. (2013) has recently presented results of new convection algorithms that are based not on mixing length theory but instead use 3D hydrodynamics as a guide to model stellar mixing.  For stellar progenitor models of SNe Ibc of masses in the range $15-27$ $\rm M_{\odot}$ they showed that enhanced convection may lead to a severe depletion of the helium layers, leaving behind only a thin helium shell comprised primarily of heavier elements.  This work suggests that the helium is not observed simply because it is not present the star at the time of explosion.  

In stars that have retained some helium before exploding, Dessart et al. (2012; D12 hereon) showed that the nickel needs to be close enough to the helium (i.e. within a $\gamma$-ray mean-free path) so it is then excited.  Thus for Ibc ejecta of similar helium abundances, ejecta where the nickel/cobalt is more thoroughly mixed will be a SN Ib, while ejecta that is weakly mixed, or strongly asymmetric (including a jet structure), will be a Ic.  

For the majority of the SNe Ib modelled here, the best-fit values from the two-zone model indicate that on average between $2-5$ times more nickel is present in the outer regions compared with the inner.  These values are reminiscent of those found for the hypernovae modelled by M03, indicating that the geometry of SNe Ib possess a degree of asymmetry.  Two of the three SNe Ic bolometric LCs were adequately described by the one-zone model, while the bolometric LC of SN 2007gr was best fit by the two-zone model, with $2-3$ times more nickel in the outer layers.  Thus for SN 2007gr the lack of helium lines cannot be accredited to poor mixing, and is likely due to the absence of helium in the ejecta.  It is then tempting to suggest that for the other two SNe Ic the lack of helium lines is due to poor mixing, possibly arising to them having a more spherically symmetric geometry.  Certainly IR helium lines were observed by Taubenberger et al. (2006) for SN 2004aw, both the He $\lambda$10,830 and $\lambda$20,580 lines, indicating that some helium is present in the ejecta.  However, the strong absorption features near $\lambda$10,830 in the spectra of SN 2011bm (Valenti et al. 2012) can be explained by C I  ($\lambda$10,827 \AA), which is also the case for SN 2007gr (Valenti et al. 2008).  Additionally, neither SN 2011bm nor SN 2007gr show evidence for a He I $\lambda$20,580 line.  Thus it appears that no helium is present in the ejecta of SN 2007gr and SN 2011bm, but \textit{is} present for SN 2004aw, implying that the latter is a Ic due to poor mixing.  Conversely, one is also drawn to conclude that should helium have been present in the ejecta of SN 2007gr, there may have been enough asphericity, and in turn sufficient mixing present, for this event to have been a type Ib.

Similarly for the majority of the SNe Ib in our sample the two-zone model was found to be a better fit to the data than the one-zone model.  This suggests that they possess an aspherical geometry, and the classification of Ib arises from the close proximity of helium and nickel in the ejecta (i.e. is thoroughly mixed).  Though we are not attempting to draw global conclusions based on a modest sample of events, our preliminary analysis implies that a large number of SNe Ib likely possess an aspherical geometry.  Studies of nebular spectra of SNe Ibc (Modjaz et al. 2008; Meada et al. 2008; Taubenberger et al. 2009), in particular the double-peaked [O I] $\lambda$6300,6363 line, have also drawn the conclusion that a very high fraction of SNe Ibc arise from aspherical explosions, with only a small fraction having lines that were best fit by a single Gaussian (as opposed to a double-peak profile, or profiles that displayed varying degrees of asphericity).  In the sample of 39 SNe Ibc investigated by Taubenberger et al. (2009), only two SNe Ib were best-fit by a single Gaussian, and for both of these events the spectra were of low S/N, making their best-fit model of these two events somewhat tentative.  It is very interesting to note that the bulk of their SNe whose lines were best fit by a single Gaussian (i.e. possess spherical symmetry) were predominantly of type Ic.

The last question that begs answering then is how can so much nickel/cobalt be in the outer layers of the ejecta?  What does this result imply for the explosion mechanism?  The classification of the type Ib of the SNe in our sample is unambiguous -- there are clear helium lines in the spectra analysed by several different authors.  As noted by D12, to ensure a observational classification of Ib, the ``right'' amount of mixing needs to occur in order that the helium in the ejecta undergoes non-thermal excitation.  This then restricts the explosion physics through the efficiency of the mixing in the outflow, where even moderately mixed models fail to non-thermally excited helium atoms.  Mixing in SN ejecta out to large radii can arises via different mechanisms which are fundamentally dependent on multidimensional effects (D12).  Large-scale mixing can occur due to a jet, large asymmetries in the explosion itself brought about by hydrodynamic instabilities, or via the magnetorotational mechanism (see D12 and references therein), while small-scale mixing can occur that is associated with convection and Rayleigh-Taylor and Kelvin-Helmholtz instabilities. The models of D12 that successfully created SNe Ib involved the explosion of low-mass helium cores in which efficient small-scale mixing occurs, and for which there is only a small oxygen-rich mass boundary between the nickel and the outer layers that are helium rich.  As most of the SNe Ib, as well as type Ic SN 2007gr, are best described by the two-zone model, which is a 1D approximation to 3D asymmetry, the amount of mixing in these events, as well as the mechanism to propel the nickel to the outer layers, can be attributed to the asymmetrical geometry of the outflow. 


Finally, while mixing is important for producing helium features in SNe Ib and a lack of these features in some SNe Ic can be attributed to a lack of extensive mixing, there are also cases (e.g. SN 2007gr) where the Ic classification is more likely to arise from the absence of helium in the ejecta. As also surmised by previous authors (e.g. Filippenko 1997), there is likely to be variation in the helium content in the progenitor's envelope, which is influenced by pre-explosion circumstances.  All massive stars lose mass during their short lifetimes, where both single and binary stars will lose mass via line-driven winds, the rate of which is highly dependent on the metal content of the star (massive stars of higher metal content lose more mass than those of lower metallicity).  This situation is also complicated by the very likely role of binarity in the progenitors stars of SNe Ibc.  Observationally there is compelling research showing that massive stars preferentially occur in binary systems (e.g. Sana et al. 2012), and it has been suggested by several authors (e.g. Podsiadlowski et al. 1992; Smartt 2009; Eldridge et al. 2013) that binaries may be the dominant progenitor route for SNe Ibc. Binary interaction offers a natural way for a star to be stripped of its outer envelope, which in addition to stellar winds, offers a way for the outer layers of the pre-explosion star to be deficient in hydrogen and/or helium. 

\section{Caveats}

Two important caveats must be considered.  The first is regarding interpreting the very late-time behaviour of the bolometric LCs, especially in those cases where only a few datapoints are available.  It is possible that SNe Ibc may have material surrounding them that was ejected by the progenitor prior to exploding, either episodically, by stellar winds or perhaps ejection during a common-envelope phase.  The time-resolved glowing circumstellar rings around SN 1987A (e.g. Panagia et al. 1991) are a prime example, which is also brilliantly illustrated in Fig. 1 of Leibundgut \& Suntzeff (2003), where the V-band ring brightness is $2-3$ magnitudes brighter than the ejecta emission.  It is not possible to spatially resolve the different emission regions for events occurring in distant galaxies, thus there may be the danger of interpreting late-time emission generated by the interaction of the SN ejecta with pre-SN ejecta, and not arising from the SN ejecta itself.  It is worth pointing out therefore that in this work we have assumed that emission observed at all times arises solely from the SN ejecta.

The second caveat concerns our analytical models.  Essentially the LC fitting consists of two regimes.  At the bolometric peak the luminosity is assumed to be equal to the instantaneous energy deposition rate, which is an approximation that arises from analytical modelling but is also seen in numerical simulations.  At later times the luminosity is also assumed to equal the instantaneous energy deposition rate but corrected for the escape of $\gamma$-rays into space.  While both fitting regions make this assumption, the physics underpinning this assumption are very different.

Deviations from the instantaneous energy deposition rate have been seen in simulations (e.g. Ensman \& Woosley 1988; Moriya et al. 2010; Bersten et al. 2012), including the recent models of Dessert et al. (2011; their Fig. 13) where the peak luminosity of their simulated SNe Ibc can exceed the instantaneous deposition by $0.1-0.3$ dex due to ``trapped'' radiation diffusing to the surface of the ejecta.  In these cases the \textit{actual} mass of nickel present in the ejecta is \textit{less} than is assumed from the instantaneous deposition rate, and any attempts to fit a simple analytical model to the bolometric peak will end up over-estimating the total amount of nickel in the ejecta.  However, deviation from the instantaneous deposition rate will be less during the exponential tail regime as the optical depth in this phase is much less.  

In our procedure we first fit every SN with the one-zone model and tried to match the peak bolometric luminosity with the tail.  In several cases we found that the model luminosity at the tail phase was under predicted (e.g. SN 2009jf, bottom left of Fig. \ref{figure:Ib}), leading us to add an additional component to the model in order for it to match the observations.  Including diffusion effects in the one-zone model will lead to a reduction in the amount of $^{56}$Ni in the model, which in turn will decrease the predicted tail luminosity.  A reduction in the tail luminosity highlights the need for an additional component in the model, but with even more $^{56}$Ni required in the outer zone to fit the observations.

The opposite effect can, in principle, also happen.  If a SN is best fit by the one-zone model but the luminosity at a given epoch exceeds the the instantaneous deposition rate, an additional component may be required to explain the observations.  However, we are not able to envision a reason why this effect should happen to just SNe Ib or Ic but not both, implying that our general conclusions are not fundamentally changed.

\section{Conclusions}
\label{section:conclusions}

\begin{itemize}
 \item There are two He I components in the optical and NIR spectra of type Ib SN 1999dn.  High-velocity He I lines are seen in both epochs considered here, and their blueshifted velocity decreases between the two epochs.
 \item Both the spectral and bolometric modelling indicate that the outflow of SN 1999dn has at least two components: an inner region of slower velocity and higher density, and an outer region of higher velocity and lower density.
 \item The bolometric LC of SN 1999dn is best described by the two-zone model, with roughly equal amounts of nickel in both regions ($\approx 0.02$ $\rm M_{\odot}$). The two-zone model allows us to approximately model 2D or 3D asymmetry and mixing with a 1D model.  Thus the likely asymmetric structure of SN 1999dn provides a physical explanation for how the radioactive material is propelled to, and mixed within, the outer regions of the ejecta.
 \item We fit the bolometric LCs of six additional SNe Ibc (three Ib and three Ic). Of the seven SNe modelled, four are best described by the two-zone model (three Ib and one Ic), two by the one-zone model (two Ic), and one is uncertain (Ib).
 \item Of the SNe Ic, only SN 2007gr was best fit by the two-zone model, indicating that for this SN Ic, the lack of helium spectral features cannot be ascribed to poor mixing.
\end{itemize}

Our approach has used observations to address the debate regarding the presence/absence of helium in the ejecta of SNe Ibc, however uncertainties still persist.  As such we encourage IR spectroscopic observations of future SNe Ibc in order to help solve this puzzle.  Detecting He I $\lambda$10,830 is not trivial as this line is often blended with other atoms and ions, however He I $\lambda$20,580 does not suffer such blending.  Detection of these lines, coupled with modelling such as that presented here can help determine whether a SNe Ic is a Ic due to poor mixing, or due to the absence of helium in the ejecta.

\begin{table*}
\scriptsize
\centering
\setlength{\tabcolsep}{3.0pt}
\caption{Best-fit parameters from the one- and two-zone models}
\label{table:params}
  \begin{tabular}{ccccccccccc}
  \hline
SN	&	Type	&	Model	&	Fitted Parameters	&	$+40$ days	&	$R^{2}$	&	$+50$ days	&	$R^{2}$	&	$+60$ days	&	$R^{2}$	&	1 or 2 zones	\\
\hline																					
1999dn	&	Ib	&	One-zone: p+t	&	$M_{\rm Ni}$ ($\rm M_{\odot}$)	&	-	&	-	&	$0.041\pm0.007$	&	0.993 (all)	&	$0.041\pm0.007$	&	0.998 (all)	&	-	\\
-	&	-	&	-	&	$\tau_{0}$	&	-	&	-	&	$4360\pm1100$	&	-	&	$6260\pm1260$	&	-	&	-	\\
-	&	-	&	One-zone: t only	&	$M_{\rm Ni}$ ($\rm M_{\odot}$)	&	-	&	-	&	$0.030\pm0.002$	&	0.748 (all)	&	$0.024\pm0.001$	&	0.603 (all)	&	\textbf{2}	\\
-	&	-	&	-	&	$\tau_{0}$	&	-	&	-	&	$18100\pm450$	&	0.956 (t)	&	$38000\pm1300$	&	1.000 (t)	&	-	\\
-	&	-	&	Two-zone:p+t	&	$M_{\rm Ni,in}$ ($\rm M_{\odot}$)	&	-	&	-	&	$0.019\pm0.000$	&	1.000 (all)	&	-	&	-	&	-	\\
-	&	-	&	-	&	$M_{\rm Ni,out}$ ($\rm M_{\odot}$)	&	-	&	-	&	$0.021\pm0.007$	&	-	&	-	&	-	&	-	\\
-	&	-	&	-	&	$\tau_{0,in}$	&	-	&	-	&	$404000\pm80000$	&	-	&	-	&	-	&	-	\\
-	&	-	&	-	&	$\tau_{0,out}$	&	-	&	-	&	$1400\pm350$	&	-	&	-	&	-	&	-	\\
-	&	-	&	Arnett	&	$M_{\rm Ni}$ ($\rm M_{\odot}$)	&	$0.040\pm0.005$	&		&	-	&	-	&	-	&	-	&	-	\\
-	&	-	&	-	&	-	&	-	&	-	&	-	&	-	&	-	&	-	&	-	\\
2007Y	&	Ib	&	One-zone: p+t	&	$M_{\rm Ni}$ ($\rm M_{\odot}$)	&	$0.042\pm0.007$	&	0.999 (all)	&	$0.042\pm0.004$	&	1.000 (all)	&	$0.039\pm0.011$	&	1.000 (all)	&	-	\\
-	&	-	&	-	&	$\tau_{0}$	&	$780\pm80$	&	-	&	$850\pm15$	&	-	&	$1700\pm900$	&	-	&	-	\\
-	&	-	&	One-zone: t only	&	$M_{\rm Ni}$ ($\rm M_{\odot}$)	&	$0.020\pm0.000$	&	0.663 (all)	&	$0.048\pm0.003$	&	0.987 (all)	&	$0.007\pm0.041$	&	0.020 (all)	&	\textbf{uncertain}	\\
-	&	-	&	-	&	$\tau_{0}$	&	$2660\pm20$	&	0.996 (t)	&	$710\pm30$	&	1.000 (t)	&	$25000\pm24350$	&	1.000 (t)	&	-	\\
-	&	-	&	Two-zone:p+t	&	$M_{\rm Ni,in}$ ($\rm M_{\odot}$)	&	$0.006\pm0.002$	&	0.999 (all)	&	$0.004\pm0.018$	&	1.000 (all)	&	$\approx0.042$	&	1.000 (all)	&	-	\\
-	&	-	&	-	&	$M_{\rm Ni,out}$ ($\rm M_{\odot}$)	&	$0.067\pm0.016$	&	-	&	$0.042\pm0.024$	&	-	&	$\approx0.006$	&	-	&	-	\\
-	&	-	&	-	&	$\tau_{0,in}$	&	$4500\pm870$	&	-	&	$3750\pm2900$	&	-	&	$\approx730$	&	-	&	-	\\
-	&	-	&	-	&	$\tau_{0,out}$	&	$210\pm10$	&	-	&	$350\pm290$	&	-	&	$\approx910$	&	-	&	-	\\
-	&	-	&	Arnett	&	$M_{\rm Ni}$ ($\rm M_{\odot}$)	&	$0.034\pm0.003$	&	-	&	-	&	-	&	-	&	-	&	-	\\
-	&	-	&	-	&	-	&	-	&	-	&	-	&	-	&	-	&	-	&	-	\\
2008D	&	Ib	&	One-zone: p+t	&	$M_{\rm Ni}$ ($\rm M_{\odot}$)	&	$\approx0.047$	&	0.989 (all)	&	$\approx0.047$	&	0.994 (all)	&	$\approx0.047$	&	0.999 (all)	&	-	\\
-	&	-	&	-	&	$\tau_{0}$	&	$\approx1900$	&	-	&	$\approx2150$	&	-	&	$\approx2460$	&	-	&	-	\\
-	&	-	&	One-zone: t only	&	$M_{\rm Ni}$ ($\rm M_{\odot}$)	&	$\approx0.031$	&	0.753 (all)	&	$\approx0.025$	&	0.558 (all)	&	$\approx0.025$	&	0.557 (all)	&	\textbf{2}	\\
-	&	-	&	-	&	$\tau_{0}$	&	$\approx4800$	&	0.971 (t)	&	$\approx7950$	&	0.992 (t)	&	$\approx7760$	&	0.982 (t)	&	-	\\
-	&	-	&	Two-zone:p+t	&	$M_{\rm Ni,in}$ ($\rm M_{\odot}$)	&	$\approx0.012$	&	1.000 (all)	&	$\approx0.017$	&	1.000 (all)	&	$\approx0.009$	&	-	&	-	\\
-	&	-	&	-	&	$M_{\rm Ni,out}$ ($\rm M_{\odot}$)	&	$\approx0.036$	&	-	&	$\approx0.089$	&	-	&	$\approx0.038$	&	-	&	-	\\
-	&	-	&	-	&	$\tau_{0,in}$	&	$\approx21000$	&	-	&	$\approx9100$	&	-	&	$\approx162000$	&	-	&	-	\\
-	&	-	&	-	&	$\tau_{0,out}$	&	$\approx1000$	&	-	&	$\approx120$	&	-	&	$\approx1450$	&	-	&	-	\\
-	&	-	&	Arnett	&	$M_{\rm Ni}$ ($\rm M_{\odot}$)	&	$0.039\pm0.001$	&	-	&	-	&	-	&	-	&	-	&	-	\\
-	&	-	&	-	&	-	&	-	&	-	&	-	&	-	&	-	&	-	&	-	\\
2009jf	&	Ib	&	One-zone: p+t	&	$M_{\rm Ni}$ ($\rm M_{\odot}$)	&	$0.166\pm0.007$	&	0.975 (all)	&	$0.165\pm0.008$	&	0.979 (all)	&	$0.163\pm0.007$	&	0.989 (all)	&	-	\\
-	&	-	&	-	&	$\tau_{0}$	&	$1640\pm20$	&	-	&	$1810\pm115$	&	-	&	$2300\pm190$	&	-	&	-	\\
-	&	-	&	One-zone: t only	&	$M_{\rm Ni}$ ($\rm M_{\odot}$)	&	$0.108\pm0.029$	&	0.812 (all)	&	$0.072\pm0.005$	&	0.479 (all)	&	$0.064\pm0.003$	&	0.387 (all)	&	\textbf{2}	\\
-	&	-	&	-	&	$\tau_{0}$	&	$3600\pm1160$	&	0.924 (t)	&	$11700\pm2200$	&	0.951 (t)	&	$17800\pm1200$	&	0.996 (t)	&	-	\\
-	&	-	&	Two-zone:p+t	&	$M_{\rm Ni,in}$ ($\rm M_{\odot}$)	&	$0.030\pm0.005$	&	0.996 (all)	&	$0.038\pm0.001$	&	1.000 (all)	&	$0.032\pm0.002$	&	-	&	-	\\
-	&	-	&	-	&	$M_{\rm Ni,out}$ ($\rm M_{\odot}$)	&	$0.151\pm0.011$	&	-	&	$0.185\pm0.018$	&	-	&	$0.516\pm0.075$	&	-	&	-	\\
-	&	-	&	-	&	$\tau_{0,in}$	&	$4e6\pm0.2e6$	&	-	&	$0.12e6\pm1.1e6$	&	-	&	$27700\pm6500$	&	-	&	-	\\
-	&	-	&	-	&	$\tau_{0,out}$	&	$940\pm130$	&	-	&	$500\pm10$	&	-	&	$120\pm65$	&	-	&	-	\\
-	&	-	&	Arnett	&	$M_{\rm Ni}$ ($\rm M_{\odot}$)	&	$0.138\pm0.010$	&	-	&	-	&	-	&	-	&	-	&	-	\\
-	&	-	&	-	&	-	&	-	&	-	&	-	&	-	&	-	&	-	&	-	\\
2004aw	&	Ic	&	One-zone: p+t	&	$M_{\rm Ni}$ ($\rm M_{\odot}$)	&	$0.080\pm0.022$	&	0.995 (all)	&	$0.080\pm0.023$	&	0.996 (all)	&	$0.081\pm0.021$	&	0.999 (all)	&	-	\\
-	&	-	&	-	&	$\tau_{0}$	&	$5100\pm1800$	&	-	&	$5100\pm1600$	&	-	&	$6050\pm2020$	&	-	&	-	\\
-	&	-	&	One-zone: t only	&	$M_{\rm Ni}$ ($\rm M_{\odot}$)	&	$0.076\pm0.014$	&	0.990 (all)	&	$0.064\pm0.008$	&	0.932 (all)	&	$0.057\pm0.007$	&	0.865 (all)	&	\textbf{1}	\\
-	&	-	&	-	&	$\tau_{0}$	&	$6150\pm1600$	&	0.967 (t)	&	$11500\pm2850$	&	0.992 (all)	&	$30400\pm16350$	&	0.996 (t)	&	-	\\
-	&	-	&	Two-zone:p+t	&	$M_{\rm Ni,in}$ ($\rm M_{\odot}$)	&	$0.032\pm0.002$	&	0.998 (all)	&	$0.045\pm0.006$	&	1.000 (all)	&	$0.052\pm0.006$	&	-	&	-	\\
-	&	-	&	-	&	$M_{\rm Ni,out}$ ($\rm M_{\odot}$)	&	$0.048\pm0.025$	&	-	&	$0.035\pm0.029$	&	-	&	$0.028\pm0.029$	&	-	&	-	\\
-	&	-	&	-	&	$\tau_{0,in}$	&	$56800\pm6000$	&	-	&	$40900\pm8440$	&	-	&	$35250\pm6000$	&	-	&	-	\\
-	&	-	&	-	&	$\tau_{0,out}$	&	$3250\pm1000$	&	-	&	$1950\pm220$	&	-	&	$820\pm400$	&	-	&	-	\\
-	&	-	&	Arnett	&	$M_{\rm Ni}$ ($\rm M_{\odot}$)	&	$0.077\pm0.011$	&	-	&	-	&	-	&	-	&	-	&	-	\\
-	&	-	&	-	&	-	&	-	&	-	&	-	&	-	&	-	&	-	&	-	\\
2007gr	&	Ic	&	One-zone: p+t	&	$M_{\rm Ni}$ ($\rm M_{\odot}$)	&	$0.029\pm0.006$	&	0.961 (all)	&	$0.029\pm0.006$	&	0.979 (all)	&	$0.029\pm0.006$	&	0.990 (all)	&	-	\\
-	&	-	&	-	&	$\tau_{0}$	&	$1600\pm300$	&	-	&	$2000\pm400$	&	-	&	$2500\pm550$	&	-	&	-	\\
-	&	-	&	One-zone: t only	&	$M_{\rm Ni}$ ($\rm M_{\odot}$)	&	$0.013\pm0.001$	&	0.570 (all)	&	$0.012\pm0.000$	&	0.510 (all)	&	$0.012\pm0.000$	&	0.491 (all)	&	\textbf{2}	\\
-	&	-	&	-	&	$\tau_{0}$	&	$13500\pm950$	&	0.975 (t)	&	$17800\pm60$	&	0.992 (t)	&	$19750\pm125$	&	0.994 (t)	&	-	\\
-	&	-	&	Two-zone:p+t	&	$M_{\rm Ni,in}$ ($\rm M_{\odot}$)	&	$0.009\pm0.001$	&	1.000 (all)	&	$0.009\pm0.002$	&	1.000 (all)	&	$0.008\pm0.001$	&	-	&	-	\\
-	&	-	&	-	&	$M_{\rm Ni,out}$ ($\rm M_{\odot}$)	&	$0.021\pm0.013$	&	-	&	$0.022\pm0.081$	&	-	&	$0.021\pm0.015$	&	-	&	-	\\
-	&	-	&	-	&	$\tau_{0,in}$	&	$26000\pm1400$	&	-	&	$24200\pm7200$	&	-	&	$28600\pm3100$	&	-	&	-	\\
-	&	-	&	-	&	$\tau_{0,out}$	&	$430\pm100$	&	-	&	$360\pm300$	&	-	&	$590\pm300$	&	-	&	-	\\
-	&	-	&	Arnett	&	$M_{\rm Ni}$ ($\rm M_{\odot}$)	&	$0.026\pm0.003$	&	-	&	-	&	-	&	-	&	-	&	-	\\
-	&	-	&	-	&	-	&	-	&	-	&	-	&	-	&	-	&	-	&	-	\\
2011bm	&	Ic	&	One-zone: p+t	&	$M_{\rm Ni}$ ($\rm M_{\odot}$)	&	$0.368\pm0.008$	&	0.991 (all)	&	$0.361\pm0.009$	&	0.987 (all)	&	$0.369\pm0.004$	&	0.991 (all)	&	-	\\
-	&	-	&	-	&	$\tau_{0}$	&	$6967\pm144$	&	-	&	$7250\pm100$	&	-	&	$7200\pm120$	&	-	&	-	\\
-	&	-	&	One-zone: t only	&	$M_{\rm Ni}$ ($\rm M_{\odot}$)	&	$0.364\pm0.008$	&	0.991 (all)	&	$0.333\pm0.007$	&	0.973 (all)	&	$0.318\pm0.002$	&	0.951 (all)	&	\textbf{1}	\\
-	&	-	&	-	&	$\tau_{0}$	&	$7150\pm175$	&	0.989 (t)	&	$8892\pm188$	&	0.987 (t)	&	$9950\pm560$	&	0.985 (t)	&	-	\\
-	&	-	&	Two-zone:p+t	&	$M_{\rm Ni,in}$ ($\rm M_{\odot}$)	&	$0.090\pm0.000$	&	0.998 (all)	&	$0.095\pm0.000$	&	0.997 (all)	&	$0.088\pm0.012$	&	0.998 (all)	&	-	\\
-	&	-	&	-	&	$M_{\rm Ni,out}$ ($\rm M_{\odot}$)	&	$0.292\pm0.009$	&	-	&	$0.281\pm0.010$	&	-	&	$0.288\pm0.003$	&	-	&	-	\\
-	&	-	&	-	&	$\tau_{0,in}$	&	$395000\pm17000$	&	-	&	$6.46e6\pm4.45e6$	&	-	&	$0.49e6\pm1.94e6$	&	-	&	-	\\
-	&	-	&	-	&	$\tau_{0,out}$	&	$4740\pm130$	&	-	&	$4760\pm100$	&	-	&	$5000\pm275$	&	-	&	-	\\
-	&	-	&	Arnett	&	$M_{\rm Ni}$ ($\rm M_{\odot}$)	&	$0.349\pm0.022$	&	-	&	-	&	-	&	-	&	-	&	-	\\

\hline

\end{tabular}

\begin{flushleft}
$^{\dagger}$The value of the correlation coefficient for the one-zone model is insensitive to the actual slope of the tail, whose precise value is necessary for determining the nickel mass.  Therefore a good statistical fit to the one-zone model doesn't imply that an accurate nickel mass has been obtained.  For the one-zone model to be deemed a good fit (e.g. SN 2004aw), the nickel mass determined from fitting the tail should be close to that from fitting the peak+tail, which is then reflected as a high $R^{2}$ value when the tail values are fit onto the peak+tail.
\end{flushleft}

\end{table*}

\section*{Acknowledgments}

We would like to thank P. Jakobsson, G. Leloudas and the anonymous referee for their excellent comments and suggestions on the original manuscript. ZC gratefully acknowledges support from a Project Grant from the Icelandic Research Fund.  SS acknowledges support from the Iniciativa Cientifica Milenio grant P10-064-F (Millennium Center for Supernova Science), with input from "Fondo de Innovaci\'{o}n para la Competitividad, del Ministerio de Econom\'{\i}a, Fomento y Turismo de Chile", and Basal-CATA (PFB-06/2007).  The work by KM is partly supported by Grant-in-Aid for Scientific Research of JSPS (23740141), and by World Premier International Research Center Initiative (WPI Initiative), MEXT, Japan.


\bsp

\label{lastpage}

\end{document}